\documentclass[11pt,twoside]{article}

\usepackage{asp2004}
\usepackage{epsf}
\usepackage{psfig}
\usepackage{lscape}
\usepackage{amssymb}

\newcommand{\kms}{\ensuremath{{\rm km\,sec^{-1}}}}                   

\newcommand{\rh}{\mbox{$r_{\rm h}$}}

\newcommand{\tdis}{\mbox{$t_{\rm dis}$}}

\newcommand{\Msun}{\mbox{$M_{\odot}$}}

\newcommand{\Msunyr}{\mbox{$M_{\odot}$~yr$^{-1}$}}

\newcommand{\be}{\mbox{\begin{equation}}}
\newcommand{\ee}{\mbox{\end{equation}}}

\newcommand{\Mmax}{\mbox{$M_{\rm max}$}}

\newcommand{\Mi}{\mbox{$M_i$}}
\newcommand{\Mp}{\mbox{$M$}}

\newcommand{\Cref}{\mbox{$m_{\rm ref}$}}

\newcommand{\qev}{\mbox{$q_{\rm ev}$}}

\newcommand{\tfour}{\mbox{$t_4$}}

\newcommand{\nbody}{\mbox{$N$-body}}

\newcommand{\Vdrift}{\mbox{$V_{\rm drift}$}}

\newcommand{\tsh}{\mbox{$t_{\rm spir}$}}

\newcommand{\tdg}{\mbox{$t_{\rm GMC}$}}
\newcommand{\tds}{\mbox{$t_{\rm spir}$}}

\newcommand{\Msunmyr}{\mbox{$M_{\odot}\,$Myr$^{-1}$}}

\begin{document}
\title{Star clusters in the solar neighborhood:\\
      a solution to Oort's problem}    
\author{Henny J.G.L.M. Lamers and Mark Gieles}   
\affil{Astronomical Institute, Utrecht University, Princetonplein 5, 3584-CC Utrecht, The Netherlands}    

\begin{abstract} 
In 1958 Jan Oort remarked that the lack of old clusters in the solar neighborhood (SN) 
implies that clusters are destroyed on a timescale of less than a Gyr.
This is much shorter than the predicted dissolution time of clusters due to stellar evolution and
two-body relaxation in the tidal field of the Galaxy.
So, other (external) effects must play a dominant role in the destruction
of star clusters in the solar neighborhood.
We recalculated the survival time of initially bound star clusters in the solar
neighborhood taking into account: (1) stellar
evolution, (2) tidal stripping, (3) perturbations by spiral arms
and (4) encounters with giant molecular clouds (GMCs). 
We find that encounters with GMCs are the most damaging to clusters.
The resulting predicted dissolution time of these combined effects,
$\tdis= 1.7 (\Mi/10^4\,\Msun)^{0.67}$
Gyr for clusters in the mass range of $10^2 < \Mi < 10^5\,\Msun$,
is very similar to the disruption time 
of  $\tdis= 1.3 \pm 0.5 (\Mi/10^4\,\Msun)^{0.62}$ Gyr
that was derived empirically from a mass limited sample of clusters in the
 solar neighborhood within 600 pc.
The predicted shape of the age distribution of clusters
agrees very well with the observed one.
The comparison between observations and theory implies 
a surface star formation rate (SFR) near the sun of   
$3.5~10^{-10}$ \Msun yr$^{-1}$pc$^{-2}$ for stars {\it in bound clusters} 
with an initial mass in the range of $10^2$ to $3~10^4$ \Msun. This can be compared to a total SFR
of 7 - 10 $\times 10^{-10}$ \Msun yr$^{-1}$pc$^{-2}$ derived from embedded 
clusters  or $3 - 7 ~10^{-9}$ \Msun yr$^{-1}$pc$^{-2}$ derived from field stars.
This implies an  infant mortality rate
of clusters in the solar neighborhood between 50\% and 95\%,
in agreement with the results of a study of embedded clusters.
\end{abstract}

\section{Introduction}

Jan Oort pointed out in 1958  
that ``in general, galactic clusters do not appear to grow much older than 0.5 Gyrs''
 (Oort 1958). 
Later, Wielen (1971) derived a mean dissolution time of 0.2
Gyr from the age distribution of open clusters within about 1 kpc from the Sun.
Spitzer (1958) suggested that the short lifetime may be due to the fact that low density 
clusters ($\lesssim 1 \Msun$pc$^{-3}$) may be destroyed by encounters with passing 
interstellar clouds.

The destruction of star clusters occurs in two stages (see Bastian \& Gieles, these proceedings, 
for a review.): \\
(1) At young ages ($<$ 10 or 20 Myr)  most of the clusters dissolve because the gas that is left over from
the giant molecular cloud (GMC) is removed from the cluster by stellar winds and supernovae.
This loss of binding energy causes a large fraction of the embedded clusters to dissolve into the field.
 This ``infant mortality'' (Lada \& Lada 2003)
 depends critically on the star formation efficiency and appears 
to be independent of the initial cluster mass. 
(Bastian et al. 2005 suggest that this occurs 
within $\sim$ 10 Myr, whereas Fall et al. 2005 suggest that it occurs over a much
longer timescale of $\sim$ Gyr). Theoretical studies, e.g. Kroupa \& Boily (2002) and 
references therein, show that the gas removal phase and the resulting infant mortality last about 
$\sim$ 10 Myr (see also Lada \& Lada 2003). \\
(2) At later ages clusters can be destroyed by external effects such as the tidal field,
and perturbations due to encounters with GMCs, with spiral arms and the Galactic disk. This effect
is mass dependent: massive clusters survive longer than low mass clusters.\\
This paper deals with the second effect, but we also derive the infant mortality rate for
clusters in the solar neighborhood (SN).

Starting with the pioneering work of Spitzer on the dynamics of star clusters,
many studies have been devoted to the understanding and prediction of the survival times of clusters
open and globular clusters in our galaxy.
These theories predicts that the dissolution time of clusters depends on their
initial mass: massive clusters survive longer than low mass
clusters (e.g. Spitzer 1958; Wielen 1985; Chernoff \& Weinberg 1990; Gnedin \& Ostriker 1997
and references therein).
Baumgardt \& Makino (2003) 
(hereafter BM03) showed from
\nbody\ simulations that the predicted dissolution time of a cluster in the tidal
field of the galaxy depends on the 
number of stars, $N$, in the cluster as $\tdis \sim \beta \left(N/\ln \gamma N\right)^x$, where 
$\gamma N$ is the Coulomb logarithm
 and the constants $\beta$ and $x$ depend on the initial concentration 
of the cluster.
Gieles et al. (2004) showed that this can be written as 

\begin{equation}
\tdis = \tfour (\Mi / 10^4 \Msun)^{\gamma}
\label{tdisformula}
\end{equation}
with $\gamma=0.62$, where \tfour\ is the disruption time of a cluster with an initial mass $\Mi=10^4\Msun$.
This same power-law dependence was
derived empirically from a study of cluster samples in four
galaxies by Boutloukos \& Lamers (2003)(hereafter BL03).
The dynamical models of BM03 result in a $\tfour = 6.9$ Gyr for clusters in the SN. 
Since the mean mass of the open clusters
in the SN is about a  $10^3$ \Msun, 
the ``mean lifetime'' of clusters in the SN, predicted by BM03, would be a few Gyr, i.e.
much longer than the empirical value of 0.2 Gyr derived by Wielen (1971).

To solve this discrepancy, we have started a series of studies of different
 effects that might play a role in limiting cluster lifetimes.
 These effects are:\\
 (1) mass loss by stellar evolution,\\
 (2) dissolution by two-body relaxation in the tidal field of the Galaxy,\\
 (3) perturbations due to the passage of spiral arms,\\
 (4) perturbations due to encounters with GMCs. \\
At the same time we rederived empirically the 
cluster dissolution times of open clusters in the SN, based on the new
cluster sample of 
Kharchenko et al. (2005). We compare the predicted and the observed
age distributions in clusters  to see which one of the possible
destruction mechanisms is the most important one in the SN. This comparison also
results in an estimate of the star formation rate and infant mortality rate.
(For the full study, see Lamers \& Gieles 2006.)

\section{The  observed age distribution of clusters in the solar neighborhood}

%
\begin{figure}
\centerline{\psfig{figure=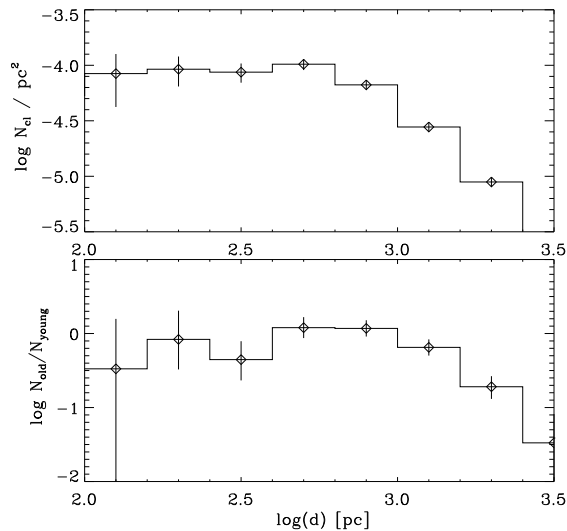,width=10.4cm}}
\caption[]{Top: The surface density distribution of open clusters in the 
Kharchenko et al. (2005) sample, projected onto the Galactic plane.
 Bottom: The ratio between the number of old ($>0.25$ Gyr) 
and young ($<0.25$ Gyr) clusters as function of distance. Both distributions are flat 
within their uncertainties up to 600 pc.
}
\label{fig:distance}
\end{figure}
%

Kharchenko et al. (2005) published a catalog of
520 galactic open clusters  in the SN with the values of angular sizes of cluster
cores and coronae, heliocentric distances $d$, $E(B-V)$, mean proper
motions, radial velocities and ages. These parameters have been
determined by homogeneous methods and algorithms including a careful procedure
of cluster member selection. The basis of this study is the ASCC-2.5
- All-Sky Compiled Catalog of about 2.5 million stars 
down to $V \simeq 14$ (completeness limit at $V \simeq 11.5$),
with proper motions and $B,V$ magnitudes based on
the $Tycho-2$ data and supplemented with $Hipparcos$ data and some ground-based catalogs.
Cluster membership is based on a combined probability which takes into
account kinematic (proper motion), photometric and spatial
selection criteria. 
Cluster ages were determined with an
isochrone-based procedure which provides a uniform age scale. This resulted in the most
homogeneous catalog of open clusters in the SN. 
Lamers et al. 2005a (L05) have shown that the lower mass limit 
of the clusters in the Kharchenko et al. sample is about 100 \Msun.

\begin{figure}
\centerline{\psfig{figure=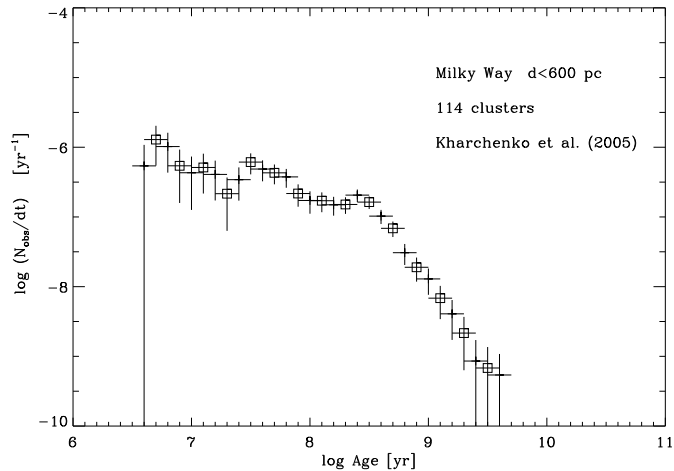,width=9.4cm}}
\caption[]{ The age histogram in units of number per
year, in logarithmic age-bins of 0.2 dex,
of 114 open clusters 
within $d<600$ pc  from Kharchenko et al. (2005). 
In order to show the effect of binning the data, the
distributions are plotted for two sets of bins, shifted by 0.1 dex,
with and without squares respectively. The error bars indicate the 
1$\sigma$ statistical uncertainty. The distribution decreases to older ages,
with a small bump around $\log (t/{\rm yr}) \simeq 8.6$. For 
$\log (t/{\rm yr}) < 7.5$ the distribution is uncertain due to large
error bars.
}
\label{fig:kharchenko_age}
\end{figure}

Figure \ref{fig:distance} shows the distance  distribution
of the density of clusters projected onto the
Galactic plane, in number per pc$^2$. 
Within the statistical uncertainty the surface density is constant 
up to at least 600 pc.
The lower part of the figure shows the ratio between old ($ t>
0.25$ Gyr) and young ($ t<0.25$ Gyr) clusters as a function of
distance. Up to a distance of about 1 kpc there is no
significant change in this ratio within the statistical uncertainty.
This is important for our study because it shows that the 
sample of clusters within 600 pc has no age bias.
Figure \ref{fig:kharchenko_age} shows the age distribution 
in number per year of the 114 clusters within 600 pc in the Kharchenko
sample. 
This distribution decreases with age, apart from a small
local maximum around $\log(t/{\rm yr}) \simeq 8.6$. The distribution
at young ages is sensitive to the choice of the age-bins and shows a
significant scatter. The steep slope at $\log(t/{\rm yr}) > 8.8$
demonstrates that cluster disruption is important.

\section{The evolution of the mass of star clusters}


\subsection{Mass loss by stellar evolution}

The stellar mass loss from clusters has been calculated by various groups.
 We adopt the $GALEV$ models for clusters with a Salpeter type
mass function in the range of $0.15 < M/\Msun<85$
(Anders \& Fritze-v. Alvensleben 2003).
These models are
based on stellar evolution tracks from the Padova group, which include
overshooting,  mass loss due to stellar winds and supernovae.
L05 have shown that the fraction of the initial cluster mass that is lost
by stellar evolution, $\qev(t)=\Delta M/\Mi$, can be approximated
accurately by

\begin{equation}
\log \qev(t)= (\log t-a_{\rm ev})^{b_{\rm ev}}+c_{\rm
  ev}~~{\rm for}~~ t>12.5~{\rm Myr} .
\label{eq:qevgot}
\end{equation}
with $t$ in yrs and $a_{\rm  ev}=7.00$, $b_{\rm ev}=0.255$ and $c_{\rm ev}=-1.805$ 
for solar metallicity models with a Salpeter mass function.

This function describes the mass loss of the models at
$t>12.5$ Myr with an accuracy of a few percent. The mass loss at 
younger ages is negligible because stars with $M_*>30~
\Msun$ hardly contribute to the mass of the cluster due to the small number of stars.
(These stars do play an important role in the infant mortality because they are responsible for the
fast removal of the gas from the clusters.)
The mass loss from clusters by stellar evolution can then be described as 

\begin{equation}
\left(\frac{dM}{dt}\right)_{\rm evol} = -M(t)\frac{d \qev}{dt}.
\label{eq:dmdtevol}
\end{equation}

\subsection{Mass loss by the galactic tidal field}

BM03 have calculated a grid of \nbody\ simulations of clusters in
circular and elliptical orbits in the tidal field of 
a galaxy for
different initial cluster masses, galactocentric distances $R$, and
different cluster density profiles. The stars follow a Kroupa initial mass function
without primordial binaries,
and stellar evolution  is taken into account during the evolution. 
BM03 and 
Lamers et al. (2005b) have shown that 
the predicted dissolution time can be expressed as a function of the initial
cluster mass as

\begin{equation}
\tdis = \tfour~(\Mi/10^4\,\Msun)^{0.62},
\label{eq:tdis}
\end{equation}
where $\tfour$ is a constant that depends on the tidal field strength
of the galaxy in which the
cluster moves and on the ellipticity of its orbit.
We adopt the value of $\tfour = 6.9\times 10^9$ yr from BM03 for clusters
in circular orbits at $R_0=8.5$ kpc. 
Applying this result of BM03 to clusters in the SN, which are on average of
lower mass than considered in their study, results in an overestimation
of the dissolution time of low mass clusters by less than a factor 2 or so 
(see Fig. 5 of BM03).
This is because low mass clusters, 
with total lifetimes less than about a Gyr,
contain massive stars during most of their lifetime. These stars are very efficient in ejecting
stars by two-body interactions. 

The mass loss due to the Galactic tidal field can then be written as

\begin{equation}
\left(\frac{dM}{dt}\right)_{\rm tidal} = \frac{-M(t)}{\tdis} = 
 \frac{-(M/10^4\Msun)^{0.38}}{\tfour / 10^4}\,\Msunyr .
\label{eq:dmdtdis}
\end{equation}

\subsection{Mass loss by spiral arm perturbations}

When clusters cross a spiral arm, the enhancement of the ambient density 
gives rise to time dependent tidal forces that can accelerate stars out of 
the cluster. These external perturbations are most 
destructive for clusters with low density and for passages with low 
relative velocity, $V_{\rm drift}$. The solar neighborhood is close to the 
corotation radius of the spiral arms, where the effect is largest. 
Although the time between spiral arm passages at that location is long,
$\propto V_{\rm drift}^{-1}$, the mass loss per passage is high, i.e. $\propto 
V_{\rm drift}^{-2}$ (Gieles et al. 2006a, hereafter GAPZ06).

GAPZ06 calculated the dissolution time
of clusters in the SN due to perturbations by spiral arms by means of N-body simulations.  
For the density of the clusters,
they adopted the observed mean mass-radius relation of
clusters as observed in nearby spiral galaxies by 
Larsen (2004):
$\rh = 3.75\ (M / 10^4 \Msun)^{\lambda}$ with $\lambda=0.10\pm0.03$, where $\rh$ is the
half mass radius. Here we have converted the cluster radius $r_{\rm eff}$ given by  
Larsen into the halfmass radius $r_{\rm h}$ using $r_{\rm eff}=0.75 r_{\rm h}$ (Spitzer 1987).
For a density contrast of the gas component of the spiral arms 
based on observations, and a mean drift velocity of
$\Vdrift=12.5$ (Dias \& L\'epine 2005), GAPZ06 found a mean dissolution time of

\begin{equation}
\tsh  =  20\,\left(\frac{M}{10^4\,\Msun}\right)\left(\frac{3.75\,\mbox{pc}}{\rh}\right)^3\,
      =  20\,\left(M/10^4\,\Msun\right)^{1-3 \lambda}\,{\rm Gyr}.
\label{eq:tdisarm2}
\end{equation}
Notice that for $\lambda = 0.1$ the disruption time due to shocking by
spiral arms has almost the same mass dependence, i.e. $\propto
M^{0.7}$, as that due to the tidal field (BM03)  and as found empirically (BL03), viz. $M^{0.62}$.  

The mass loss of clusters due to spiral arm shocks is 

\begin{equation}
\left(\frac{dM}{dt}\right)_{\rm spir} =  \frac{-M(t)}{\tds} 
 = -0.5 \left(\frac{M(t)}{10^4\,\Msun}\right)^{0.3}  ~ \Msunmyr.
\label{eq:dmdtarm1}
\end{equation}

\subsection{Mass loss by giant molecular cloud encounters}

Spitzer (1958) suspected already that encounters with GMCs play an important role in the destruction 
of clusters in the SN. Similar to the case of spiral arm perturbations, the disruptive effect of the 
encounters depends on the density of the star cluster and on the relative velocity between the cluster 
and the cloud. However in the case of an encounter with a GMC, the mass loss also depends on 
the mass of the GMC and on the distance of the passage at the moment of closest approach. 
This implies of course that the dissolution of clusters due to encounters with GMCs can only be 
described in a statistical way, when all effects are properly averaged 
over the encounter probabilities. 

Gieles et al. (2006b, GPZB06) studied the encounters between GMCs and clusters with $N$-body
simulations. 
They showed that the relative mass loss due to a shock, i.e. $\Delta M/M$,
 is only about 20 \% of the 
relative energy gain $\Delta E/E$. 
This is because a large fraction ($\sim$ 80 \%) of the energy gained
in an encounter goes into ejecting stars with a velocity exceeding the escape velocity.
This implies that the disruption time should {\it not} be defined as the timescale for injecting the
same energy as the cluster binding energy (e.g. Ostriker et al. 1972),
 but as the time scale to bring the cluster mass to zero.

The disruption time of clusters by encounters with GMCs does not
depend on the mass of the individual clouds, but only on the {\it mean
  density of molecular gas} in the SN. This is because the energy
gained by the cluster per individual encounter is proportional to the
cloud mass $M_{\rm cloud}$, but the number of encounters per unit time
depends on the number density of the clouds. The number density times the mass of
the clouds is the mean density of the molecular gas. In other words:
1000 clouds of $10^4~\Msun$ have statistically the same disruptive effect as 10
clouds of $10^6~\Msun$ if we ignore the tidal field (but see \S 3.5).

GPZB06 derived an expression for the energy gain and the
resulting mass loss for the full range of encounter distances, from
head-on to distant encounters. 
They adopted
a mean midplane density of molecular gas in GMCs  near the sun of
$\rho_{\rm n}=0.03 \Msun$pc$^{-3}$, a surface density of individual GMCs of
$\Sigma_{\rm n}=170 \Msun$ pc$^{-2}$ 
(Solomon et al. 1987) and a mean
velocity dispersion of clusters and GMCs of $\sigma_v \simeq 10~\kms$.

With these data  they derived a mean
dissolution time ($\tdg$) for clusters by GMC encounters in the SN, 
taking into account the relative velocity distribution, the distribution
of impact parameters and gravitational focusing.
 Adopting the mean mass radius relation of clusters (Larsen 2004) (see \S 3.3) they found
  
\begin{equation}
\tdg =  2.0\,\left(\frac{M}{10^4\,\Msun}\right)\left(\frac{3.75 \mbox{ pc}}{\rh}\right)^3
     =  2.0\,\left(M/10^4\,\Msun\right)^{0.7}\,{\rm Gyr},
\label{eq:tdisgmc1}
\end{equation}
The statistical mass loss rate of a cluster due to encounters with GMCs is 

\begin{equation}
\left(\frac{dM}{dt}\right)_{\rm GMC} = \frac{-M(t)}{\tdg} =  
-5.0\,\left(\frac{M(t)}{10^4\,\Msun}\right)^{0.3} \Msunmyr.
\label{eq:dmdtgmc1}
\end{equation}
Notice that the mass dependence is the same as for dissolution by
spiral arm shocking, but that the effect is ten times stronger.

\subsection{The predicted mass evolution of clusters in the solar
neighborhood}
\label{subsec:masslosspredictions}

%
\begin{figure}
\centerline{\psfig{figure=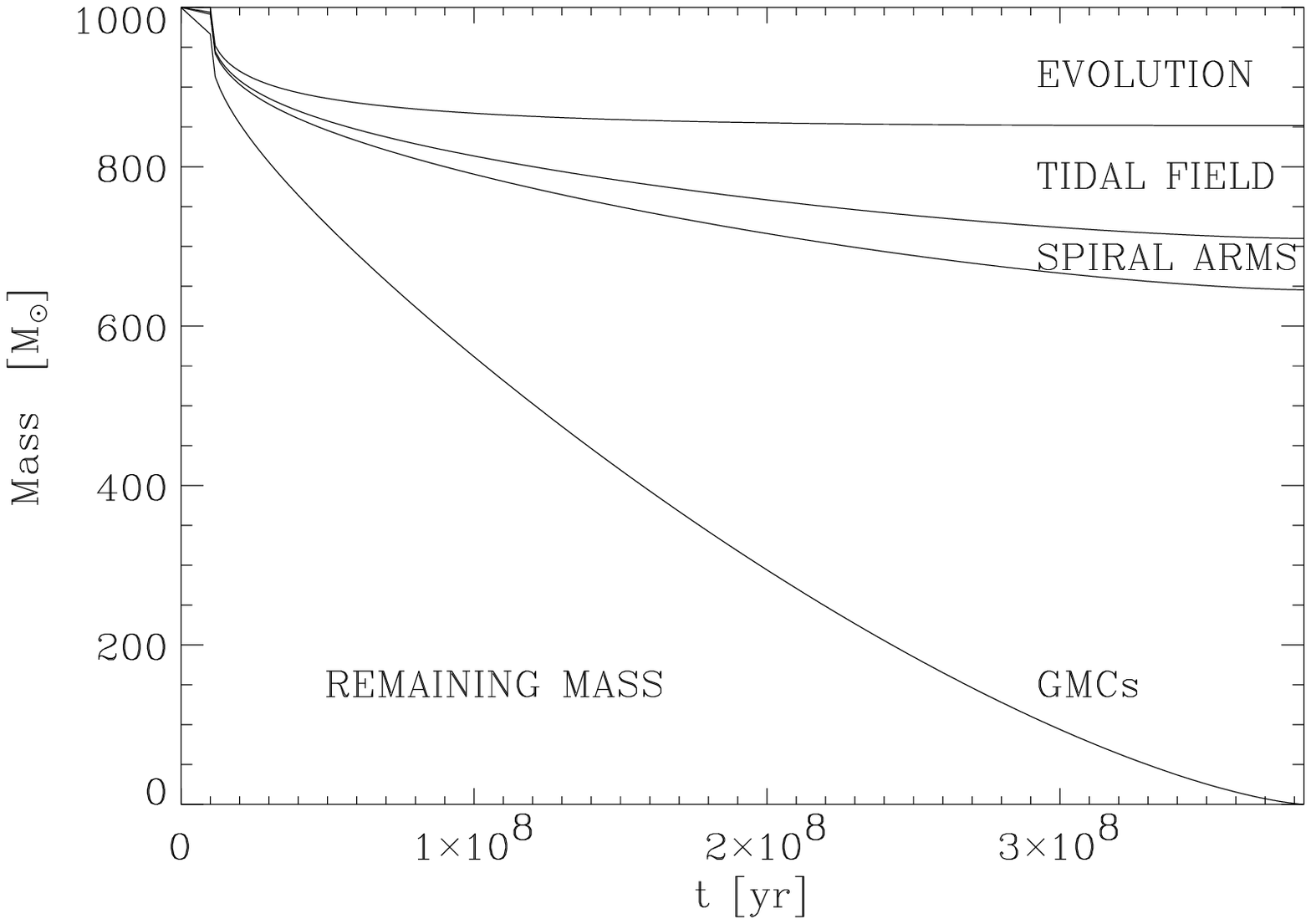,width=7.0cm}
            \psfig{figure=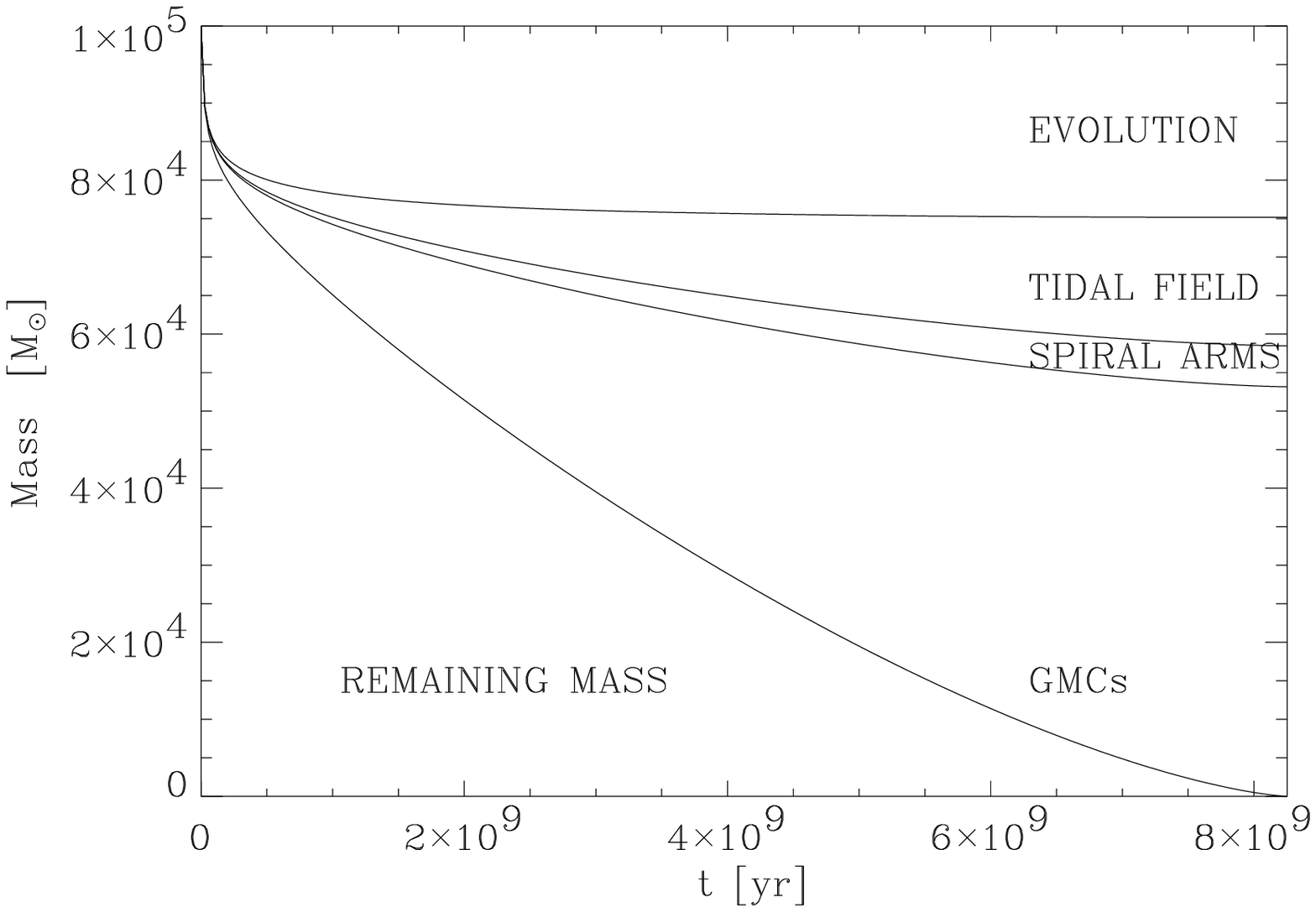,width=7.0cm}}
\caption{The mass  evolution of a cluster with an initial mass of
  $10^3$\, (left) and $10^5$ \Msun\ (right) in the solar neighborhood.
The mass loss due to the four separate effects
is indicated. Encounters with GMCs
are the dominant dissolution effect in the solar neighborhood. A cluster of $M_i=10^3$ \Msun\ 
dissolves in 0.35 Gyrs and a cluster of $10^5$ \Msun\ in 8.2 Gyrs.}
 \label{fig:5}
\end{figure}
%

The mass loss from clusters,
due to the combined effects of stellar evolution, tidal stripping, spiral arm 
shocks and encounters with GMCs, is

\begin{equation}
\frac{d\Mp}{dt}= \left(\frac{d\Mp}{dt}\right)_{\rm evol} +
                 \left(\frac{d\Mp}{dt}\right)_{\rm tidal}+
                 \left(\frac{d\Mp}{dt}\right)_{\rm spir} +
                 \left(\frac{d\Mp}{dt}\right)_{\rm GMC}.
\label{eq:dmpdt}
\end{equation}
 In writing this expression we assumed
that the four effects act independently of one another. This may not be the case.
For instance, a cluster that has encountered a GMC has an increased radius
 and is thus more susceptible to tidal stripping
than a cluster that did not have a recent encounter. However, we expect 
this effect to be small, except for the lowest mass clusters.

\begin{figure}
\centerline{\psfig{figure=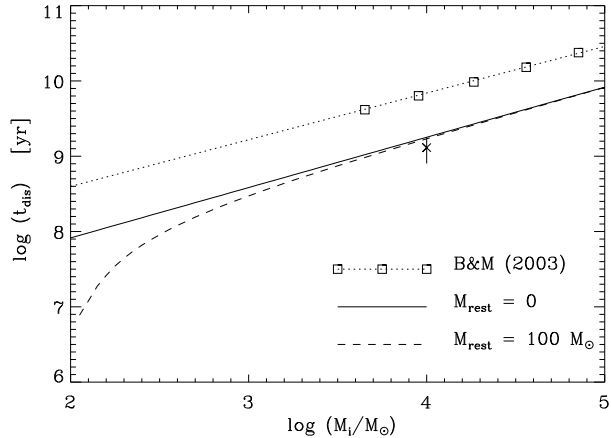,width=9.0cm}}
\caption[]{The predicted dissolution times of clusters in the solar
neighborhood due to the combined effects of stellar evolution, 
tidal field, spiral arm
shocks and encounters with GMCs, as a function of the initial mass.
 Full line: total dissolution time. Dashed line: time when the
 remaining mass is 100 \Msun. Squares and dotted line: dissolution time due to
 stellar evolution and the Galactic tidal field only, predicted by
 BM03. 
 Cross with error bar: the value of $t_4$ empirically derived by L05.}
\label{fig:6}
\end{figure}

 We have solved Eq. \ref{eq:dmpdt}
numerically for clusters of different masses. The results are shown in
Fig. \ref{fig:5} for clusters with an initial mass of $10^3$ and $10^5$ 
\Msun. Notice that encounters with GMCs are the dominant dissolution 
effect in the solar neighborhood, 
contributing as much as the three other effects combined. 
Using the calculated mass decrease of clusters of different initial mass
we can predict the age of the cluster when it has a remaining mass of 100 \Msun,
i.e. the detection limit for clusters in the SN in the Kharchenko sample (see L05).
Figure \ref{fig:6} shows the ages of clusters when their remaining
mass is 0 and 100 \Msun\ as a function of the initial mass. 
The almost linear part can be described by 

\begin{equation}
\tdis=1.7 (\Mi/10^4~\Msun)^{0.67}~{\rm Gyr ~~~~~~~~~~~~ for}~~~~ 3.5 \simeq \log{\Mi/\Msun} \simeq 5
\end{equation}
 The steep decrease at low mass is because clusters with $\Mi < 10^3$ \Msun\ 
quickly reach the detection limit of $100\,\Msun$.
The figure also shows the dissolution times by the Galactic 
tidal field, predicted by
BM03 for clusters with an initial concentration factor $W_0=5$
in a circular orbit at $R_0=8.5$ kpc and the empirical value of \tfour\ from L05.
 Our predicted timescales are about a factor 5 smaller than those due
the tidal field only (BM03) but agree with the empirical value (L05). 
%

\section{Comparison with observed age distribution of clusters
in the solar neighborhood}

Given the initial mass distribution of the clusters, their formation
rate, CFR$(t)$, and the time it takes for a dissolving cluster to fade
below the detection limit, we can predict
the distribution of observable clusters as function of age
or mass\footnote{
L05 have derived an expression for the general 
case of a cluster sample that is set by a {\it magnitude limit} for any formation history}. 
Here we are interested in a sample of clusters with $M>10^2~\Msun$. 

\begin{figure}
\centerline{\psfig{figure=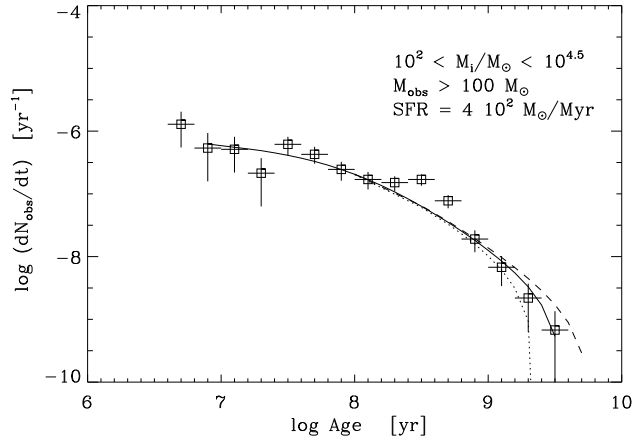,width=9.0cm}}
\caption[]{The observed age distribution of an unbiased sample of
clusters with $M>100\, \Msun$ in the solar neighborhood within 600 pc
(Karchenko et al. 2005; L05) in units of nr~yr$^{-1}$ is given by
squares with the Poisson error bars. The full line shows the predicted
distribution for a cluster sample with a maximum mass of $M_{\rm
max}=3~10^4~\Msun$ and a SFR of $4~10^2\,\Msunmyr$ within 600 pc from
the Sun.  The dotted and dashed lines are for $M_{\rm
max}=1.5~10^4~\Msun$ and $M_{\rm max}=6~10^4~\Msun$ respectively. The
bump around 0.4 Gyr suggests an increased SFR at that time.  }
\label{fig:7}
\end{figure}

For a constant CFR and a power law cluster IMF with
a slope of $-\alpha=-2$ 
(Lada \& Lada 2003) the number of clusters with
$M>100~\Msun$ as function of age is

\begin{equation}
N_{M>100}(t) = C ~ ( M_{\rm lim}(t)^{-1} - M_{\rm max}^{-1} ),
\label{eq:Nt}
\end{equation}
where $M_{\rm lim}(t)$ is the {\it initial} mass of clusters that
reach $M(t)=100\,\Msun$ at age $t$.  Clusters of age $t$ with a smaller initial
mass have $M<100\,\Msun$ by now.  $\Mmax$ is the maximum {\it initial}
mass of the clusters that are formed. The constant $C$ is related to the star
formation rate (SFR) in bound clusters as 
${\rm SFR}=C \ln (M_{\rm max}/M_{\rm lim})$ for $-\alpha=-2$.

Figure \ref{fig:7} shows a comparison between the observed age distribution of clusters
within 600 pc  
with the predicted distribution for
$\Mmax=3~10^4 \Msun$. This value of \Mmax\ is adopted because the
observed distribution shows a steep drop at $\log~t \simeq 9.5$  (with
only one cluster in the last bin) and
Fig. \ref{fig:7} shows that this corresponds to $\Mi = 3~10^4\,\Msun$.

The flattening of the predicted distribution at the low age end is due to the
fact that clusters with an initial mass in the range of
about 100 to 300 \Msun\ quickly reach 100
\Msun\ (see Fig. \ref{fig:6}). The bump in the observed distribution
around $\log (t)\simeq 8.6$ is due to a local starburst (see L05 and 
Piskunov et al. 2006).
There is good agreement in the shapes of the predicted and observed
distributions!

The vertical shift that is applied to the predicted curve
to match the observed one gives a value of $C=10^{-4.15}$ in
Eq. \ref{eq:Nt},
which corresponds to a SFR of $4~10^2\,\Msunmyr$ for bound clusters in the 
range of 
$10^2 < \Mi/\Msun < 3~10^4$ within 600 pc from the sun and a surface formation rate of 
 $3.5~10^{-10}$ \Msun yr$^{-1}$pc$^{-2}$. 
We can derive the infant mortality rate in the SN by comparing this value with the
total starformation rate of at least  $7 - 10 ~10^{-10}$ \Msun yr$^{-1}$pc$^{-2}$, 
derived from embedded clusters with a mass $\>$ 35 \Msun, by Lada \& Lada (2003)
or the rate of $3 - 7 ~10^{-9}$ \Msun yr$^{-1}$pc$^{-2}$, derived from the field star population
by Miller \& Scalo (1979). 
Our value is a factor 2 to 3 smaller than the lower limit of Lada \& Lada and a factor
10 to 20 smaller than the one derived by Miller \& Scalo.
This implies  that at least $\sim$70 and possibly even 0.95 \% of the stars in the SN 
are born in clusters that dissolve within $\sim$ 10 Myr.

\section{Conclusions and discussion}

 Our calculated dissolution times of clusters in the solar
neighborhood are about a factor five smaller than predicted by BM03 
for clusters in the tidal field of our Galaxy, with only stellar evolution,
binaries and two-body relaxation taken into account. This is mainly due to
encounters with GMCs. So we can expect that clusters in environments with
a high density of GMCs will be destroyed very effectively. This is especially
 the case in interacting galaxies, such as M51 and the Antennae
galaxies. Indeed,  
Gieles et al. (2005)
derived a disruption time
\tfour\ of only 0.2 Gyr for clusters in M51 at a Galactocentric distance of $1 < R < 5$ kpc.
This is more than a factor 10 smaller than expected on the basis of the tidal field only.
On the other hand, in galaxies with a small GMC density the dissolution time will be set mainly by
the tidal field. This is confirmed by the comparison between empirical and predicted
dissolution times of several galaxies by Lamers et al. (2005b) 
(see Fig. 6 of Bastian \& Gieles, these proceedings).


\section*{Acknowledgments}

We thank the organizers of this very nice and stimulating workshop. 
We thank Nina Kharchenko and her colleagues for discussions about their catalogue of
galactic clusters and Soeren Larsen for comments on the mass-radius dependence of
clusters. We are grateful to Holger Baumgardt, Douglas Heggie and Roland Wielen for usefull
 comments on this study.
This work is supported by a grant from the Netherlands Research
School for Astronomy (NOVA).




\section*{Note added after submission}

In a recent paper 
Whitmore, Fall \& Chandar (astro-ph 0611055)
 argue that our determination
of the disruption time of clusters in the solar neighborhood  is
influenced by selection effects.
They claim that ``not including the young embedded clusters from Lada \&
Lada (2003) can produce an apparent bend in the age distribution, which
can be misinterpreted in the BL03 models as evidence for a specific
value of the disruption time''. 
This is a curious statement because the embedded clusters  are all
younger than 3 Myr and the large majority of these clusters
will not survive infant mortality (Lada \& Lada 2003). 
Therefore they should not be included
in a study of the dissolution of {\it bound clusters}.

\begin{figure}[!h]
\centerline{\psfig{figure=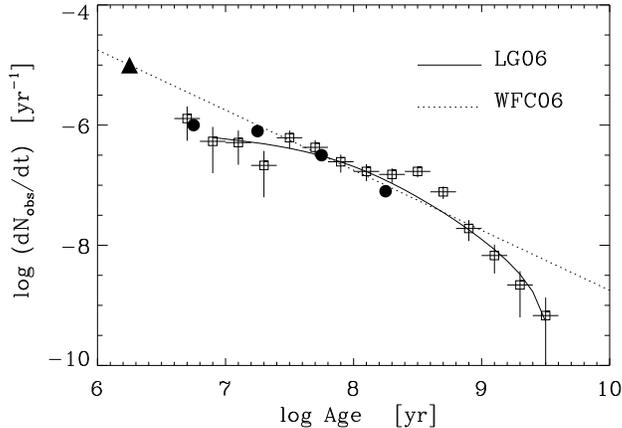,width=9.0cm}}
\caption[]{Open squares: the age distribution of clusters in the SN (our
sample).  Filled symbols: the age distribution adopted by Whitmore et
al. (2006).  Filled triangle: the embedded clusters of Lada \& Lada (2003) 
at $\log(t/{\rm yr})<6.5$. 
Filled circles: the open clusters by Battinelli \& Capuzzo-Dolceta (1991).
The two distributions agree in the range
of overlap.  
Full line: our fit to the data of non-embedded clusters that was used to derive the 
dissolution time of clusters that survived infant mortality. Dashed line:
Whitmore's linear fit with slope of -1, shifted vertically to match the
data at $\log(t/{\rm yr})=7.75$. This fit does not agree with the flattening of the
observed distribution at $\log(t/{\rm yr}) < 7.5$ and neglects the 
decrease at $\log(t/{\rm yr}) > 9$.  }
\label{fig:8}
\end{figure} 

Figure \ref{fig:8} compares our age distribution (open squares with error bars) 
with the one adopted by Whitmore et al.  (filled symbols). They adopted an
 age distribution 
which consists of the sample of embedded
clusters of Lada \& Lada (i.e. with ages $\lesssim 3\,$Myr) and open clusters
(i.e. which have removed their gas) from Battinelli \& Capuzzo-Dolceta 
(1991, MNRAS 249, 76).   Lada \& Lada (2003)
combined these to show the steep drop (1 dex in log(d$N$/d$t$) going
from the embedded to the unembedded phase (i.e. from $\lesssim3\,$Myr
to $\gtrsim5\,$Myr), which they refer to as ``infant mortality'' within 10 Myr.

We see that:\\
a.) the age distribution of the non-embedded clusters from  Lada \& Lada (2003)
 agrees  with the one we derived from the Kharchenko et al. (2005)
sample  in the range $6.75 <\log(t/{\rm yr})< 8.25$.\\ 
b.) the one high point at $\log(t/{\rm yr})=6.25$ is due to
embedded clusters, most of which will dissolve by infant mortality 
(Lada \& Lada 2003).\\ 
c.) the straight line adopted by Whitmore et al. is a very
poor fit to the observed age distribution as it seriously overpredicts 
the distribution at $6.7 < \log(t/{\rm yr})<7.5$ and does not explain
the drop in the number of clusters at $\log(t/{\rm yr})>8.0$.
 (The peak
in the age distribution around $\log(t/{\rm yr})\simeq 8.6$, probably due
to an increased star formation episode (L05), is higher than both fits.)\\

Therefore, we conclude that the
statement by Whitmore et al. (2006), that our derived short disruption
time is due to the neglect of the embedded clusters, is incorrect. 
In fact, Fig. \ref{fig:8} shows that the linear fit of slope -1,
proposed by Whitmore et al. in support of their model
for a mass independent infant mortality over an extended period of time,
$\sim$1 Gyr, fits the data of the non-embedded clusters poorly. Our fit
of the distribution of non-embedded clusters agrees much better with the 
observations and results in a dissolution time of bound clusters
that agrees very well with the predictions described in this paper.

\end{document}